\documentclass[a4paper,11pt]{article}
\usepackage{pos}
\usepackage{booktabs,enumitem,titlesec,tikz}
\usepackage{setspace}
\usepackage[hang,flushmargin]{footmisc}
\usepackage{microtype}

\makeatletter
\def\ps@emptyfoot{%
  \def\@oddfoot{}%
  \def\@evenfoot{}%
}
\makeatother

\makeatletter
\renewcommand\section{\@startsection{section}{1}{\z@}%
    {2.1ex \@plus 1ex \@minus .5ex}
    {0.9ex \@plus .3ex \@minus .3ex}
    {\normalfont\large\secstyle}}
\renewcommand\subsection{\@startsection{subsection}{2}{\z@}%
    {2.1ex\@plus 1ex \@minus .5ex}
    {0.4ex \@plus .2ex \@minus .2ex}
    {\normalfont\normalsize\secstyle}}
\makeatother

\let\OLDthebibliography\thebibliography
\renewcommand\thebibliography[1]{
  \OLDthebibliography{#1}
  \setlength{\parskip}{0pt}
  \setlength{\itemsep}{7pt plus 0.3ex}
}

\usepackage{tocloft}
\let\latexd\d
\newcommand{\vol}{\mathcal{V}}
\newcommand{\cloop}{c_{\rm loop}}
\newcommand{\gs}{g_s}
\newcommand{\Kahler}{K\"ahler}

\def\SM{{\scriptscriptstyle SM}}

\title{Loop Blow-up Inflation: An Overview}

\author*{Suk\latexd{r}ti Bansal}

\affiliation{Institute for Theoretical Physics, TU Wien, Wiedner Hauptstraße 8-10/136, A-1040 Vienna, Austria}

\emailAdd{sukrti.b@gmail.com}

\abstract{This proceedings contribution provides an overview of Loop Blow-up Inflation and updates its observational predictions and their comparison with the latest CMB and BAO data. It is based on work originally published in \cite{Bansal:2024uzr}, carried out in collaboration with L.~Brunelli, M.~Cicoli, A.~Hebecker, and R.~Kuespert, and presented at the 2025 Workshop on Quantum Gravity and Strings. We focus on string loop corrections to the Kähler potential, long regarded as a potential threat to blow-up inflation in the Large Volume Scenario. We argue that these corrections, previously assumed avoidable, are in fact generically present and qualitatively alter the original non-perturbative picture: they invalidate slow-roll near the minimum and instead generate a new slow-roll regime at larger field values, where the scalar potential transitions from an exponential to a power-law plateau. This leads to modified inflationary dynamics and distinct cosmological predictions, including an increased tensor-to-scalar ratio.\vspace{4pt}

For the scenario with vanishing extra dark radiation, we compare with the latest combined analyses of ACT, DESI, SPT, BICEP/Keck, and Planck, while for scenarios with non-zero extra dark radiation we use ACT DR6 constraints. Tighter observational bounds on extra dark radiation require updated model parameters in one scenario, yielding revised predictions presented here for the first time. All predictions remain consistent with recent observations, with the ACT+DESI combination yielding near-perfect agreement in the spectral index for vanishing extra dark radiation, with a deviation of $0.03\sigma$. We also comment on subleading loop corrections, which improve robustness by reducing the field value required for slow roll. These results highlight that string loop effects, rather than being merely detrimental, can play a constructive role in realising viable inflation in string compactifications.

}

\FullConference{2025 Workshop on Quantum Gravity and Strings\\
Corfu Summer Institute\\
Mon-Repos, Corfu, Greece\\
{\small 7 - 14 September 2025}
}

\tableofcontents
\begin{document}

\maketitle
\raggedbottom

\vspace*{-10pt}
\enlargethispage{1\baselineskip}
\section{Introduction}

Cosmic inflation is one of the most successful paradigms in modern cosmology. It elegantly accounts for the observed flatness and homogeneity of the universe, the near scale-invariance of the primordial power spectrum, and the absence of primordial monopoles. Yet any realistic inflationary model must ultimately be embedded in a UV-complete theory of quantum gravity. String theory provides such a framework, but its extra-dimensional compactifications generically produce a large number of light scalar fields, the moduli, whose stabilisation is a prerequisite for both theoretical consistency and phenomenological viability.
 
A key question is whether stabilising all moduli permits the flat directions needed for slow-roll inflation. Type IIB flux compactifications provide a particularly well-studied framework in which such issues can be addressed in a controlled manner. In this context, background fluxes stabilise the complex structure moduli and the axio-dilaton, while the Kähler moduli remain unfixed at tree level due to the no-scale structure of the effective supergravity. Subleading perturbative and non-perturbative effects then generate a potential for the Kähler moduli.

The Large Volume Scenario (LVS) exploits this structure to stabilise the compactification manifold's overall volume, the lightest modulus, at exponentially large values. It allows appropriate uplift to an almost-Minkowski vacuum while leaving room for an appropriate inflationary trajectory. A key observation is that due to approximate rescaling symmetries of all the moduli orthogonal to the overall volume, certain directions in Kähler moduli space remain flat at leading order. These flat directions provide natural inflaton candidates and form the basis of Kähler moduli inflation.

A prominent example is blow-up inflation, in which a local blow-up Kähler modulus acts as the inflaton and acquires an exponentially flat potential from non-perturbative effects. However, it has long been argued that string loop corrections spoil this flatness and render the model unviable. In this work, based on the analysis in \cite{Bansal:2024uzr}, we revisit this expectation and show that string loop corrections are not only unavoidable but qualitatively change the picture. While they indeed destroy slow roll in the original non-perturbative blow-up inflation regime ~\cite{Conlon:2005jm}, at larger field values they give rise to a new inflationary phase. In this regime, the potential takes a power-law plateau form generated by loop effects, leading to a novel class of models which we term Loop Blow-up Inflation.

\enlargethispage{1\baselineskip}

\section{Kähler Moduli in Type IIB Flux Compactifications}

\subsection{Geometric Setup and Moduli}

\begin{figure}[h!]
\centering
    \vspace{-1 em}
    \includegraphics[scale=0.36]{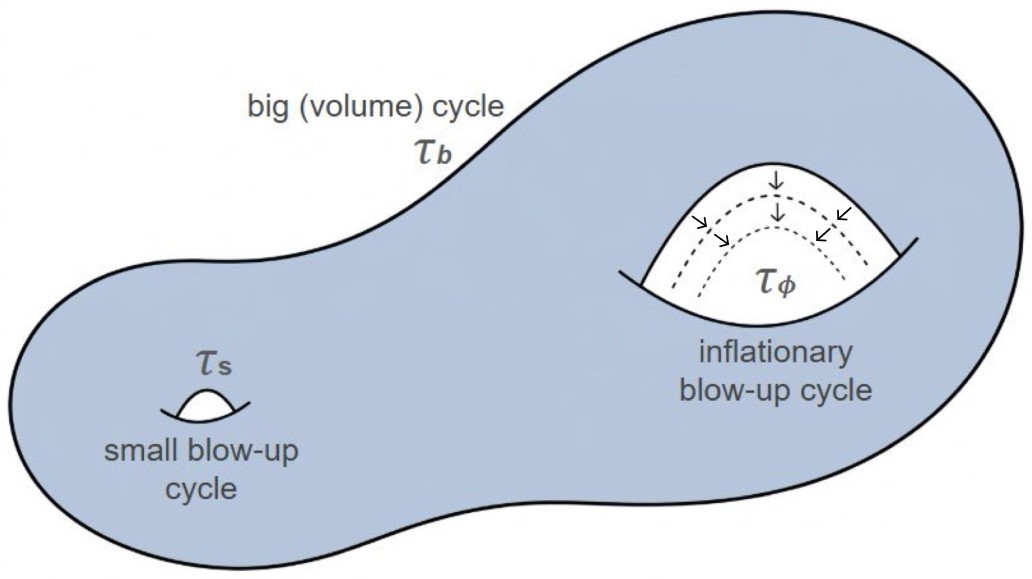}
        \vspace{-0.4 em}
    \caption{Calabi-Yau manifold with four-cycle volumes in the minimalistic model of blow-up inflation. $\tau_\phi$ shrinks during inflation (inward arrows) as it slowly rolls towards the minimum of the potential, while $\tau_s$ and $\tau_b$ remain stabilised.}
    \label{manifold}
    \vspace{-12pt}
\end{figure}

We consider a minimalistic model with three Kähler moduli $T_i$, the minimum number required for one modulus to serve as the inflaton ($T_\phi$) and another one as a small cycle ($T_s$) while keeping the overall volume fixed:\vspace{-4pt}
\begin{equation}
T_i = \tau_i + i\,c_i\,,\quad i\in\{b,\,\phi,\, s\}.
\end{equation}
Here $\tau_i$ are the real parts (four-cycle volumes, cf. Fig. \ref{manifold}) and $c_i$ are their axionic partners. The Calabi-Yau volume takes the Swiss-cheese form:\vspace{-4pt}
\begin{equation}
\vol = \tau_b^{3/2} - \lambda_\phi \tau_\phi^{3/2} - \lambda_s \tau_s^{3/2}\,,\vspace{-4pt}
\end{equation}
with $\tau_b \gg \tau_\phi, \tau_s$ as the \looseness=-1 holes must be much smaller than the cheese. The moduli $\tau_\phi$ and $\tau_s$ are the blow-up modes, while the big cycle $\tau_b$ controls the overall volume. $\lambda_s$ and $\lambda_\phi$ represent ratios of triple intersection numbers. As the Kähler moduli $T_i$ arise from the compactification geometry, they correspond to closed-string modes.

\enlargethispage{1\baselineskip}

\subsection{$N$=$1$ Supergravity Framework}\vspace*{4pt}

The four-dimensional $N$=$1$ supergravity is characterised by the superpotential \vspace{-6pt}\begin{equation}\label{eq:Superpotential}
W = W_0 + A_s e^{-a_s T_s} + A_\phi e^{-a_\phi T_\phi}, 
\vspace{-6pt}
\end{equation}
where $W_0$ is the flux-generated constant contribution and the exponential terms arise from E3-branes  or gaugino condensation. The prefactors $A_{s}$ and $A_\phi$ are determined by the complex structure moduli \newpage \noindent and are of order one. The Kähler potential is\vspace{-12pt}
\begin{equation}\label{eq:KaehlerPotential} 
K = K_{\rm cs} - 2 \ln\left(\mathcal V + \frac{\hat\xi}{2}\right),\vspace{-8pt}
\end{equation} where $K_{\rm cs}$ depends only on the complex structure moduli and the axio-dilaton, both of which are stabilised by fluxes. Therefore it plays no role in the low-energy dynamics and can be considered constant. $\hat\xi$ encodes the leading $\alpha'^3$ correction.

The F-term scalar potential is given by\vspace{-6pt}
\begin{equation}
V_F = e^K \left( K^{I \bar J} D_I W D_{\bar J} \overline W - 3 |W|^2 \right), \label{eq:FtermPotential} \vspace{-8pt}
\end{equation}
with $D_I W = \partial_I W + (\partial_I K) W$ and $D_{\bar J} \overline W$ its complex conjugate. A non-generic feature of the potential is that it has the \textit{no-scale} structure where the terms coming from the tree-level Kähler potential and consisting of the Kähler moduli, cancel the negative term $3|W|^2$, leaving a non-negative potential~\cite{Giddings:2001yu}. The naively dominant $1/\vol^2$ terms cancel, so the leading-order Kähler moduli scalar potential is generated by $\mathcal{O}(\alpha'^3)$ effects and scales as $V \sim |W_0|^2/\vol^3$. Due to the no-scale structure, the scalar potential is suppressed by inverse powers of the volume.

\subsection{Large Volume Scenario}

In the Large Volume Scenario (LVS)~\cite{Balasubramanian:2005zx,Conlon:2005ki} $\vol \gg 1$, and we already have $\tau_b \gg \tau_i$ (for $i = \phi, s$) from the Swiss-cheese form. In this regime the scalar potential \eqref{eq:FtermPotential} becomes:\vspace{-6pt}
\begin{equation}
V_{\text{LVS}} = \hat{V}\left[\sum_{i=s,\phi}\mathcal{A}_i\, \frac{\sqrt{\tau_i} \,e^{-2 a_i \tau_i}}{\vol} - \sum_{i=s,\phi}\mathcal{B}_i \, \frac{\tau_i \, e^{-a_i\tau_i}}{\vol^2} + \frac{3 \hat{\xi}}{4\vol^3}\right]\,,
\end{equation}
where\vspace{-4pt}
\begin{equation}
\hat{V} \equiv \frac{\gs e^{K_{\rm cs}}}{8\pi}\,W_0^2\,,\quad
\mathcal{A}_i \equiv \frac{8\,(a_i A_i)^2}{3{W_0^2}\lambda_i}\,,\quad
\mathcal{B}_i \equiv 4 \frac{a_i |A_i|}{W_0}\,.
\end{equation}
Large \looseness=-1 volume limit: On stabilising (minimising) $V_{\text{LVS}}$ with respect to the small cycle $\tau_s$, one finds that as\vspace{-6pt}
\allowdisplaybreaks
\begin{equation}
\allowdisplaybreaks
    \vol \rightarrow \infty,\ \  a_s\tau_s \approx \ln \vol.\vspace{-2pt}
\end{equation}
The inflaton instead, being displaced from the LVS minimum before inflation, takes larger values, giving the full hierarchy $\tau_b \gg \tau_\phi \gg \tau_s$ (cf. Fig. \ref{manifold}). $V_{\text{LVS}}$ has an AdS minimum. In order to describe the universe in the present epoch it needs to be uplifted to an approximately Minkowski minimum. For this we add a positive term\vspace{-6pt}
\allowdisplaybreaks
\begin{equation}\label{eq:uplift-def}
\allowdisplaybreaks
V_{\text{up}}(\vol) = \frac{\hat{V}\mathcal{D}}{\vol^2}\,,\vspace{-4pt}
\end{equation}
where $\mathcal{D}$ is a constant\footnote{For a precise determination of $\mathcal{D}$, see eq.~(4.7) of \cite{Cicoli:2016olq}, which fixes it so that the LVS+uplift minimum is Minkowski.}, so that the non-perturbative potential $V_\text{np}$ becomes\vspace{-6pt}
\allowdisplaybreaks
\begin{equation}\label{V_np}
\allowdisplaybreaks
    V_\text{np}=V_{LVS}+V_\text{up}\,.\vspace{-6pt}
\end{equation}
A natural question is what physical phenomena can cause the uplift. The widely known uplift mechanism of anti-D3 brane has been shown to have control issues \cite{Junghans:2022exo,Gao:2022fdi,Junghans:2022kxg,Hebecker:2022zme}. On checking its feasibility for the specific case of Loop Blow-up Inflation, we found that it is not workable for the potential with the leading-order loop correction studied in this work. However, it may become viable when subleading loop corrections are included.

\enlargethispage{1\baselineskip}
Alternative mechanisms such as D-term effects \cite{Braun:2015pza}, dilaton-dependent non-perturbative contributions \cite{Cicoli:2012fh, Retolaza:2015nvh}, T-branes \cite{Cicoli:2015ylx}, or non-zero F-terms of the complex structure moduli \cite{Gallego:2017dvd,Hebecker:2020ejb,Krippendorf:2023idy}, could be at play. Looking at the power of $\vol$ in the expression for $V_{\text{up}}(\vol)$, i.e. $(-2)$, D-term effects seem to be the most probable cause\footnote{The author is grateful to Ignatios Antoniadis for pointing this out.} for uplift.

\enlargethispage{1\baselineskip}
\section{String Loop Corrections}

\subsection{Form of the Loop Correction}\label{corr_form}

We briefly review the structure of string loop corrections. The loop corrections to the Kähler potential are generally written as the sum of Kaluza-Klein (KK) and winding (W) corrections:\vspace{-3pt}
\begin{equation}
\delta K_{(\gs)} = \delta K^\text{KK}_{(\gs)} + \delta K^\text{W}_{(\gs)}\,,\vspace{-5pt}
\end{equation}
where, schematically,\vspace{-8pt}
\begin{equation}
\delta K^\text{KK}_{(\gs)} \simeq \sum_{i} C_i^\text{KK}\,\frac{\gs \mathcal{T}^i(t^a)}{\vol}\,,\qquad
\delta K^\text{W}_{(\gs)} \simeq \sum_{i} C_i^\text{W}\,\frac{1}{\mathcal{I}^i(t^a)\vol}\,.\vspace{-7pt}
\end{equation}
These corrections were originally calculated for the torus-orbifold and then later extrapolated to generic Calabi-Yau manifold in~\cite{Berg:2007wt}. The coefficients $C_i^\text{KK}$ and $C_i^\text{W}$ are unknown functions of the complex structure moduli and are expected to be suppressed by $\pi$ factors. The functions ${\cal T}^i$ and ${\cal I}^i$ are homogeneous functions of the two-cycle volumes $t^a$ of degree 1. 

{\setlength{\parindent}{16pt}Using the field-theory interpretation of string loop corrections developed in~\cite{Cicoli:2007xp,Gao:2022uop}, building on the explicit torus-orbifold computation of~\cite{Berg:2005ja}, the loop correction to the scalar potential is found to be}\vspace{-3pt}
\begin{equation}\label{deltaVloop}
\delta V_{\text{loop}} \simeq -\,\frac{\hat{V}}{\vol^3}\,\frac{\cloop}{\vol^{1/3}}\,f\left(\frac{\vol^{2/3}}{\tau_\phi}\right)\,, \quad\text{where}\quad \cloop \simeq \begin{cases} C_i^\text{W} \\ (\gs\, C_i^\text{KK})^2 \end{cases}.\vspace{-3pt}
\end{equation}
$\delta V_\text{loop}^\text{KK}$ has an `extended no-scale structure'. $f$ encodes information from the unknown functions ${\cal T}^i$ and ${\cal I}^i$; its precise functional form in an explicit Calabi-Yau setting is unknown. When the blow-up cycle $\tau_\phi$ is smaller than any other nearby cycle, EFT arguments imply a loop correction that depends only on $\tau_\phi$ and, via the Weyl rescaling of the $4D$ metric, on ${\cal V}$. Following the estimate for open-string loops in \cite{Cicoli:2007xp} and the derivation for closed-string loops in~\cite{Gao:2022uop}, we get\vspace{-5pt}
\begin{equation}\label{eq:leading-f}
f \simeq \frac{\vol^{1/3}}{\sqrt{\tau_\phi}}\,,\vspace{-5pt}
\end{equation}
which gives us\vspace{-6pt}
\begin{equation}\label{deltaV_loop}
    \delta V_{\text{loop}} \simeq -\,\frac{\hat{V}}{\vol^3}\,\frac{\cloop}{\sqrt{\tau_\phi}}\,.\vspace{-2pt}
\end{equation}
This induces a negative contribution to the scalar potential that grows at small $\tau_\phi$. Any unknown ${\cal O}(1)$ factors in $f$ are absorbed into the coefficient $c_\textrm{loop}$. The numerical factor $\cloop$ does not involve $g_s$. From the explicit torus orbifold results of \cite{Berg:2005ja}, reference \cite{Gao:2022uop} derived the value $1/(2\pi)^4$. Alternatively, identifying the relevant cutoff with the Kaluza-Klein scale $M_p/(\tau_\varphi^{1/4}\sqrt{\mathcal{V}})$ gives the standard 4D loop factor $1/(16\pi^2)$.

\enlargethispage{1\baselineskip}

\subsection{The Full Potential}

The complete scalar potential is the sum of the LVS potential and the uplift term, eq. \eqref{V_np}, and the string loop correction \eqref{deltaV_loop}$\,$:\vspace{-5pt}
\begin{equation}
V = V_{\text{LVS}} + V_{\text{up}} + \delta V_{\text{loop}}\,.\vspace{-4pt}
\end{equation}
After stabilizing $\vol$ and $\tau_s$ in the large volume limit, and assuming the condition $\lambda_\phi a_\phi^{-3/2} \ll \lambda_s a_s^{-3/2}$ so that backreaction of the inflaton on volume stabilisation is negligible, the potential as a function of $\tau_\phi$ becomes:\vspace{-10pt}
\begin{equation}\label{eq:tauphi-pot}
V(\tau_\phi) = V_0 \left[1 + \mathcal{A}_\phi\frac{\vol^2}{\beta}\,\sqrt{\tau_\phi}\,e^{-2 a_\phi \tau_\phi} - \mathcal{B}_\phi\frac{\vol}{\beta}\,\tau_\phi\,e^{-a_\phi\tau_\phi} - \frac{\cloop}{\beta\sqrt{\tau_\phi}}\right].\vspace{-6pt}
\end{equation}
Here\vspace{-7pt}
\allowdisplaybreaks
\begin{equation}\label{eq:V0}
\allowdisplaybreaks
V_0 \equiv \left[V_\text{LVS}(\vol, \tau_s) + V_\text{up}(\vol)\right] \Big|_\text{minimum}
= \frac{\hat V\beta}{\vol^3},\vspace{-3pt}
\end{equation}
where $\beta$ as determined in Sec. 4.1 of \cite{Cicoli:2016olq} is given by\vspace{-8pt}
\allowdisplaybreaks
\begin{equation}\label{eq:beta-def}
\allowdisplaybreaks
\beta \simeq \frac32 a_\phi^{-3/2} \lambda_\phi \left(\ln \vol\right)^{3/2}.\vspace{-6pt}
\end{equation}
$\beta$ is an $\mathcal{O}(1)$ parameter. It fixes the uplifting term \eqref{eq:uplift-def} so that $V_0$ cancels the negative contribution from the two exponential terms in \eqref{eq:tauphi-pot} after minimizing in 
$\tau_\phi$. The value of $\beta$ is shifted by the $\cloop$ term, but this correction is negligible at our level of precision.

We define the canonically normalised inflaton $\phi$ in terms of the four-cycle $\tau_\phi$:\vspace{-8pt}
\begin{equation}
\label{eq:canonically normalized inflaton}
    \phi = \sqrt{\frac{4 \lambda_\phi}{3{\vol}}} ~\tau_\phi^{3/4}\,.\vspace{-2pt}
\end{equation}

\enlargethispage{1\baselineskip}
\subsection{Inevitability of String Loop Corrections}\label{inev_corr}\vspace{2pt}

\noindent\textbf{String loop corrections and the $\eta$-problem in blow-up inflation}: The original blow-up inflation model~\cite{Conlon:2005jm} relied purely on non-perturbative effects to generate a slow-roll potential. String loop corrections were believed to ruin this behaviour by inducing large contributions to the second slow-roll parameter $\eta$, violating $|\eta| \ll 1$, even after extended no-scale cancellations \cite{Cicoli:2008gp,Baumann:2014nda}.\vspace{1pt}

\enlargethispage{1\baselineskip}
\noindent\textbf{Proposed circumventions}: To avoid this problem and yet have viable physical models, two circumventions were proposed:\vspace{-0.8em}
\begin{itemize}
[itemsep=-0.3em,leftmargin=1.6em]
\item[(i)] One proposed circumvention was to consider blow-up inflation models in which no branes wrap the blow-up four-cycles, so that string-loop corrections are supposedly absent.
\item[(ii)] Additionally, it was argued that even in blow-up inflation models where string-loop corrections cannot be completely avoided, they can be made negligible by tuning the string coupling $g_s$ and the superpotential $W_0$ to be sufficiently small.
\end{itemize}\vspace{-0.7em}

\noindent\textbf{Do string loop corrections invalidate non-perturbative blow-up inflation?} We carried out a quantitative analysis of the effects of string loop corrections in Sec. 4.2 of \cite{Bansal:2024uzr}. Denoting the loop-induced correction to the slow-roll parameters $\epsilon$ and $\eta$ by $\delta\epsilon$ and $\delta\eta$ respectively, we find $\delta\epsilon/\epsilon\gg\delta\eta/\eta$. Hence the most stringent condition for the original model's predictions to remain valid is $\delta\epsilon \ll \epsilon$, which translates to $c_{\text{loop}} \ll 0.4 \times 10^{-6}$. However, the explicit torus-orbifold computation of~\cite{Berg:2005ja}, as analysed in~\cite{Gao:2022uop}, gives $c_{\rm loop} \sim (1/2\pi)^4$ and $4D$ effective field theory (EFT) logic tells us that $c_{\rm loop} \sim 1/16\pi^2$. Both these values exceed the threshold value $c_{\text{loop}} \ll 0.4 \times 10^{-6}$ by orders of magnitude. So string loop corrections are not negligible and do indeed invalidate the original blow-up inflation model.

Moreover, the EFT value $c_{\rm loop} \sim 1/16\pi^2$ also exceeds the separate threshold $c_{\rm loop} \gtrsim 6 \times 10^{-4}$ above which $|\eta| > 1$ in the original slow-roll field range. The non-perturbative contribution to $\eta$ is exponentially small there and so $|\eta|\approx |\delta\eta| \simeq 11 > 1$, confirming the $\eta$-problem suspected in the earlier literature. The worldsheet value $c_{\rm loop} \sim (1/2\pi)^4$ sits at the boundary of this threshold.

\noindent\textbf{Can loop corrections be avoided by proposed circumventions?} Our analysis reveals that the proposed circumventions are ineffective.\vspace{-0.7em}
\begin{enumerate}[itemsep=-0.3em,leftmargin=1.6em]
\item[(i)] The absence of branes wrapping del Pezzo divisors eliminates open string loops but closed string loops remain inevitable. This is because realizing the non-perturbative minimum that stabilizes $\tau_\phi$ after inflation requires a superpotential correction $\sim \exp(-a_\phi T_\phi)$ which implies the presence of an O-plane near the blow-up cycle to locally break supersymmetry to $N$=$1$. This leads to the presence of closed-string loop corrections, as discussed in~\cite{Gao:2022uop}.
\item[(ii)] The smallness of $g_s$ is constrained by the fact that the volume scales exponentially with $1/g_s$. In addition, tuning $W_0$ to small values tends to reduce the volume. This causes problems due to the  resulting warping corrections, as discussed in detail in \cite{Junghans:2022exo, Gao:2022fdi} and in Sec. 3.3 of \cite{Bansal:2024uzr}.
\end{enumerate}\vspace{-8pt}
We therefore conclude that string loop corrections are both unavoidable and large enough to invalidate the original blow-up inflation regime.

\enlargethispage{1\baselineskip}
\subsection{$\cloop$ cases}

\begin{figure}[h!]
\centering
\vspace{-5pt}
    \includegraphics[scale=0.48]{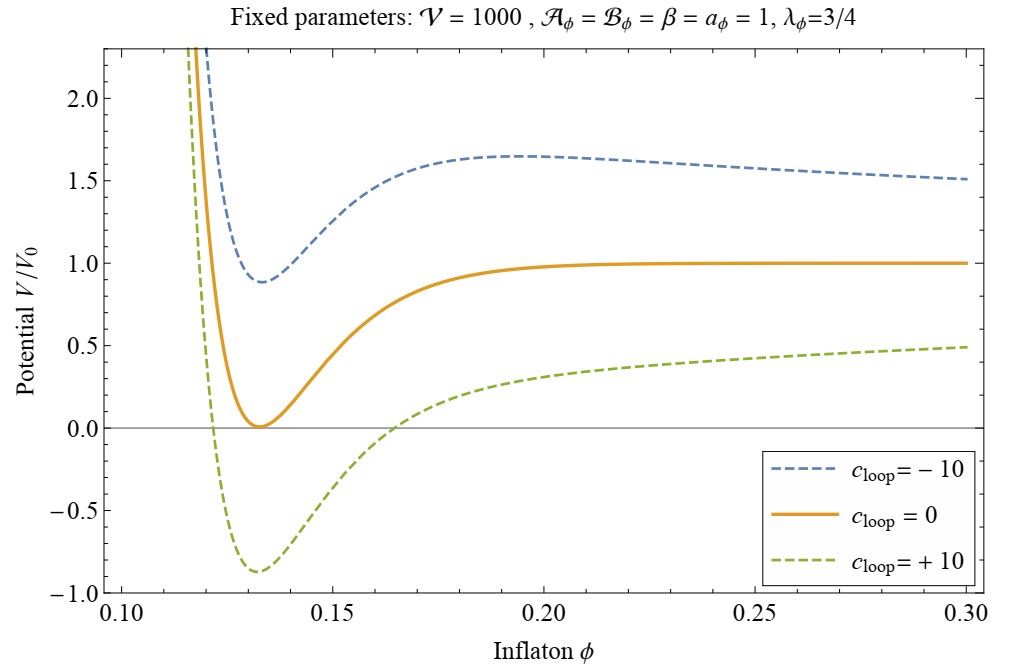}
    \vspace{-5pt}
    \caption{Plot of the potential \eqref{eq:tauphi-pot} for $\cloop=\pm10$ and $\cloop=0$.}
    \label{fig:potential}\vspace{-6pt}
\end{figure}

Now we look at plots of the potential versus the value of the inflaton. We consider three cases with $\cloop$ negative, $0$ and positive. In Fig. \ref{fig:potential}, the magnitudes of $\cloop$ values have been taken to be unrealistically large for better visibility. To match the present universe, the post-inflationary minimum must be a (near-)Minkowski vacuum after adjusting the constant term. However, in Fig. \ref{fig:potential}, we do not impose this tuning for non-zero values of $\cloop$, and for $\cloop=0$ we make the minimum exactly $0$, in order to keep the three plots sufficiently far apart from each other for better resolution.

$\cloop < 0$: The dashed blue curve shows that the potential is never flat enough to allow slow roll.

$\cloop =0$: The solid yellow curve corresponds to the original non-perturbative model of blow-up inflation. Its flat region, corresponding to slow-roll, starts relatively close to the minimum as the potential approaches a constant exponentially fast.

$\cloop > 0$: The dashed olive-green curve shows that the flatness of the potential is spoiled in the original region of slow-roll. This is consistent with the discussion in Sec. \ref{inev_corr} which tells us that $\cloop \gtrsim 10^{-6}$ destroys non-perturbative blow-up inflation. However, this plot also reveals that for such $\cloop$, slow-roll is regained at larger values of $\phi$. This shows us that slow-roll inflation is possible even in the presence of string-loop corrections. This overturns the earlier expectation that loop corrections are purely detrimental, showing instead that they can generate a viable inflationary regime. A complementary summary can be found in \cite{Brunelli:2025hcn}.

\enlargethispage{1\baselineskip}
\section{Inflationary Dynamics}

\subsection{The Inflationary Regime}\label{inf_regime}

For $\cloop\gtrsim 10^{-6}$, the regime in which the original model is physically unviable, we zoom in on that part of the potential where inflation takes place, i.e. where $\phi$ is sufficiently large so that the exponential terms in eq. \eqref{eq:tauphi-pot} can be neglected. In this region the potential can be written as \vspace{-5pt}
\begin{equation}\label{eq:potential}
V(\phi) = V_0\left(1 - \frac{b\,\cloop}{\phi^{2/3}}\right)\vspace{-5pt}
\end{equation}
where $b \equiv \sigma_\phi/(\beta \vol^{1/3})$ with $\sigma_\phi \equiv (4\lambda_\phi/3)^{1/3}$. This power-law potential represents a qualitatively \newpage \noindent new form compared to the exponential potentials characteristic of other Kähler moduli inflation models. Although $\phi$ is required to take a large value for slow roll to take place, its largeness is constrained by the hierarchy of the Large Volume Scenario which requires $\tau_\phi < \tau_b$. It has been shown via Kähler cone computations in \cite{Bansal:2024uzr} that if $\tau_\phi \sim \tau_b$ then $ \phi \sim 1$, leading to the requirement $\phi < 1$. All our inflationary analyses stay within this regime.

\subsection{Slow-Roll Parameters}

The slow-roll parameters following from the potential~\eqref{eq:potential} are:\vspace{-6pt}
\begin{equation}
\epsilon = \frac{1}{2}\left(\frac{V_\phi}{V}\right)^2 \simeq \frac{2}{9}\,\frac{(b\,\cloop)^2}{\phi^{10/3}}\,,
\end{equation}
\begin{equation}
\eta = \frac{V_{\phi\phi}}{V} \simeq -\,\frac{10}{9}\,\frac{b\,\cloop}{\phi^{8/3}}\,.\vspace{-3pt}
\end{equation}
Slow-roll is naturally achieved for small values of $(b\,\cloop)$.

\enlargethispage{1\baselineskip}

\subsection{Cosmological Observables}

The duration of inflation, quantified by the number of e-foldings $N_e$, is:\vspace{-6pt}
\begin{equation}\label{eq:e-foldings}
N_e = \int_{\phi_{\rm end}}^{\phi_*} \frac{V}{V_\phi}\,d\phi \simeq \frac{9}{16}\,\frac{\phi_*^{8/3}}{b\,\cloop}\,,\vspace{-6pt}
\end{equation}
where $\phi_{\rm end}$ is the \looseness=-1 value of $\phi$ at the end of inflation and $\phi_*$ at the scale of horizon exit. Here $\phi_{\rm end} \ll \phi_*$. The amplitude of primordial density perturbations $\hat{A}_s$ (corresponding to $12\pi^2 \tilde{A}_s\simeq 2.1 \times 10^{-9}$) given by\vspace{-10pt}
\begin{equation}\label{eq:pdp}
    \hat{A}_s=\left. \frac{V^3}{V_\phi^2}\right\rvert_{\phi=\phi_*}=\ \frac{9V_0}{4} \frac{\phi_*^{10/3}}{(b\,\cloop)^2}\,, \vspace{-4pt}
\end{equation}
takes the value $ \simeq 2.5 \times 10^{-7}$ as per CMB data~\cite{Tristram:2023haj,AtacamaCosmologyTelescope:2025blo}. In computing the above expression we have used potential \eqref{eq:potential} and in the end the approximation $(1-\cloop\, b\, \phi_*^{-2/3})\simeq 1$.

Solving eq. \eqref{eq:e-foldings} for $\vol$ in terms of $\phi_*$ gives
\vspace{-10pt}
\begin{equation}
\label{eq:volume-sol-simple}
\allowdisplaybreaks
\vol=\frac{A}{\phi^{8}_*}\,,\quad \text{where}\ A \equiv \left( \frac{16 N_e \sigma_\phi \cloop}{9 \beta} \right)^3.\vspace{-6pt}
\end{equation}
\enlargethispage{1\baselineskip}
On plugging this expression for $\vol$ into eq. \eqref{eq:pdp} we get\vspace{-10pt}
\begin{equation}
\label{eq:phi-sol-simple}
\allowdisplaybreaks
 \phi_*= \left( B A^7 \right)^{\frac{1}{66}} ~,\quad \text{where}\ B \equiv \left( \frac{4 \hat{A}_s \sigma_\phi^2 \cloop^2}{9 \beta^3 \hat{V}} \right)^3.\vspace{-6pt}
\end{equation}
Thus $\vol$ and $\phi_{*}$ are expressed in terms of $A$ and $B$. For the natural parameter choice\vspace{-6pt}
\begin{equation}\label{Parameters}
\allowdisplaybreaks
\lambda_\phi=1\,,\qquad c_{\rm loop}= 1/(16\pi^2)\,,\quad \text{and} \quad \beta=W_0=g_s\,e^{K_{\rm cs}}=2\,,
\vspace{-6pt}
\end{equation}
eqs. \eqref{eq:volume-sol-simple} and \eqref{eq:phi-sol-simple} give us\vspace{-10pt}
\begin{equation}\label{vol_exp}
\vol \simeq 1743\,N_e^{5/11} \,,\quad \text{and} \quad \phi_* \simeq 0.06\,N_e^{7/22}.\vspace{-6pt}
\end{equation}
For $N_e \simeq 52$, this gives $\phi_* \simeq 0.2$, comfortably satisfying the Kähler cone constraint $\phi < 1$ discussed in sec. \ref{inf_regime}. The spectral index $n_s$, and tensor-to-scalar ratio $r$, are:\vspace{-6pt}
\begin{equation}\label{ns_exp}
    n_s = 1+ 2 \,\eta - 6\, \epsilon \simeq 1 -\frac{20}{9}\, \frac{b\, \cloop}{\phi_*^{8/3}}\ \ \Rightarrow\ n_s \simeq 1-\frac{1.25}{N_e}\,, \vspace{-8pt}
\end{equation}
\begin{equation}\label{r_exp}
    r= 16\,\epsilon \simeq \frac{32}{9}\, \frac{(b\, \cloop)^2}{ \phi_*^{10/3}}\qquad\qquad\  \Rightarrow\ r \simeq \frac{0.004}{N_e^{15/11}}\,, \vspace{-2pt}
\end{equation}
where the right-hand-side expressions are obtained for the parameter values chosen in \eqref{Parameters}. These \newpage \noindent relations combine to give the relation:\vspace{-12pt}
\begin{equation}
r \simeq 0.003\,(1-n_s)^{15/11}.
\end{equation}

\enlargethispage{1\baselineskip}
\section{Post-Inflationary Evolution}

Post-inflationary evolution is richer than in standard single-field scenarios due to multiple light moduli whose sequential decays produce a non-standard thermal history. We have seen that inflation is driven by a single modulus, the inflaton, while the volume modulus appears as a stabilised parameter in the potential. After inflation, both the inflaton $\phi$ and the volume modulus $\chi$ (the canonically normalised volume modulus) are displaced from their post-inflationary minima and oscillate coherently, with the inflaton carrying the dominant energy density. The post-inflationary dynamics is therefore governed by both $\phi$ and $\chi$, while the remaining moduli $\tau_s$ and $\tau_{\rm SM}$ never dominate the energy density and can be neglected.

\subsection{Reheating}\label{reheating}

After inflation ends, the transfer of inflationary energy to the Standard Model begins via a process known as reheating. In LVS models, this process begins with coherent oscillations of the inflaton. Although such oscillations can, in principle, trigger non-perturbative preheating via parametric resonance, in the present setup all inflaton couplings are suppressed by inverse powers of $\vol$ or $M_p$. As a result, any resonance lies deep in the narrow regime and is inefficient, so reheating proceeds entirely through perturbative decays.

The inflaton decays first, producing radiation and possibly lighter moduli. However, the volume modulus $\chi$, whose oscillations redshift as matter, can eventually dominate over the radiation produced by the inflaton decay. This leads to a period of moduli domination, followed by the decay of $\chi$, which reheats the universe again. The resulting cosmological history is therefore non-standard, featuring successive epochs of matter and radiation domination driven by the dynamics of multiple moduli.

The precise decay channels depend on the location of the Standard Model (SM) in the compactification. Since the SM cannot reside on D7-branes wrapping the inflaton cycle (chiral intersections force the non-perturbative prefactor to vanish), it is realised either on D7-branes wrapping a separate divisor $\tau_{\rm SM}$ or on D3-branes at a singularity. Depending on whether the inflaton cycle is wrapped by a hidden D7-stack, this leads to three distinct scenarios:\vspace{-3pt}
\begin{itemize}[itemsep=-0.3em,leftmargin=1.1em]
    \item SM on D7-branes
    \begin{itemize}[itemsep=-0.2em]
    \vspace{-0.7em}
        \item[I)] Inflaton-cycle wrapped by D7s
        \item[II)] Inflaton-cycle not wrapped by D7s
    \end{itemize}\vspace{-0.3em}
    \item SM on D3-branes
    \begin{itemize}[itemsep=-0.2em]
    \vspace{-0.7em}
        \item[IIIa)] Inflaton-cycle wrapped by D7s
        \item[IIIb)] Inflaton-cycle not wrapped by D7s
    \end{itemize}
\end{itemize}

\vspace{-6pt}
$N_e$ depends on the details of this post-inflationary evolution. In particular, it is sensitive to the durations of the two moduli-dominated epochs, denoted by $N_\phi$ and $N_\chi$. One finds\vspace{-6pt}
\begin{equation}
\label{eq:N_e from reheating}
    N_e \simeq 57 +\frac{1}{4}\ln r -\frac{1}{4}\left(N_\phi + N_\chi\right) + \frac{1}{4} \ln \left(\frac{\rho_*}{\rho(t_{\rm end})}\right),\vspace{-6pt}
\end{equation}
where $\rho_*$ is the energy density at horizon exit and $\rho(t_{\rm end})$ that at the end of inflation. Given the flatness of the inflationary plateau, we take $\rho_* \simeq \rho(t_{\rm end})$, allowing us to neglect the last term.

Substituting the expression for $r$ from eq.~\eqref{r_exp} into eq.~\eqref{eq:N_e from reheating} gives\vspace{-6pt}
\begin{equation}\label{Ne_exp}
    N_e \simeq 55.6 -\frac{15}{44}\ln N_e -\frac{1}{4}(N_\phi + N_\chi)\,.\vspace{-6pt}
\end{equation}
The quantities $N_\phi$ and $N_\chi$ can be expressed in terms of the volume $\vol$ and microscopic parameters such as $W_0$, $\beta$, and $Z$ (see sec.~5.2 of \cite{Bansal:2024uzr}). Using eq.~\eqref{vol_exp} to rewrite $\vol$ in terms of $N_e$, and adopting \newpage \noindent natural values for these parameters, some of which were already chosen in eq. \eqref{Parameters}, eq.  \eqref{Ne_exp} becomes a self-consistency equation in $N_e$ alone, which can be solved numerically.

The solution gives $N_e \simeq 51.6 - 53$ depending on the SM location scenario. Once $N_e$ is determined, $n_s$ and $r$ follow directly from eqs. \eqref{ns_exp} - \eqref{r_exp}. The results are collected in Table \ref{pheno-table} and compared with observations in sec. \ref{obs}.

\enlargethispage{1\baselineskip}
\subsection{Dark radiation}

The decays of $\phi$ and $\chi$ produce not only SM particles but also ultra-light axions -- the bulk closed-string axion $a_b$ and, when the SM is on D7-branes, the SM axion $a_{SM}$. Being extremely weakly coupled, these axions remain relativistic and contribute to the effective number of additional neutrino-like species $\Delta N_{\rm eff}$. If the last modulus to decay is denoted $\sigma$, this contribution is determined by the ratio of its decay rate into hidden-sector degrees of freedom, denoted by $\Gamma_{\sigma \to {\rm hid}}$, to its decay rate into SM particles, denoted by $\Gamma_{\sigma \to \SM}$:\vspace{-5pt}
\begin{equation}
\Delta N_{\rm eff} = \frac{43}{7}\frac{\Gamma_{\sigma \to {\rm hid}}}{\Gamma_{\sigma \to \SM}}\left(\frac{10.75}{g_*(T_{\rm rh})}\right)^{1/3}   
\label{DNeff}\vspace{-3pt}
\end{equation}
with $g_*(T_{\rm rh})$ the effective number of relativistic degrees of freedom at temperature $T_{\rm rh}$ \cite{Cicoli:2012aq, Higaki:2012ar}. In Scenario I, the volume modulus decays predominantly into SM Higgs scalars, giving $\Delta N_{\rm eff}\simeq 0$. In Scenario II, a detailed accounting of all inflaton decay channels yields $\Delta N_{\rm eff} \simeq 0.14$. In Scenario III, $\Delta N_{\rm eff} \simeq 1.43/Z^2$, where $Z$ is the Giudice-Masiero coefficient controlling the coupling of the volume modulus to the Higgs doublets. The latest ACT DR6 data \cite{AtacamaCosmologyTelescope:2025nti} require $Z \gtrsim 2.9$, higher than the $Z \gtrsim 1.7$ required by the earlier Planck bound used in \cite{Bansal:2024uzr}. For $Z = 4$ one obtains $\Delta N_{\rm eff} \simeq 0.09$.

\enlargethispage{1\baselineskip}
\subsection{Observational predictions and constraints}\label{obs}

The post-inflationary history derived above must be consistent with observations from two key epochs in the later thermal evolution of the universe.\vspace*{2pt}

\textbf{Light-element abundances} \textit{(constrain $T_{\rm rh}$)}: During Big Bang nucleosynthesis (BBN), the light elements \({}^4\)He, D, \({}^3\)He and \({}^7\)Li are produced in the first few minutes after the Big Bang at temperatures \(T\sim\mathcal{O}(1)\,\mathrm{MeV}\). Their observed primordial abundances agree with standard BBN predictions only if the universe is already radiation-dominated and thermally equilibrated when nucleosynthesis begins. This requires the reheating temperature to satisfy $T_{\rm rh}\gtrsim \mathcal{O}(1)\,\mathrm{MeV}$. The values of \(T_{\rm rh}\) obtained in our three scenarios lie between \(10^8\) and \(10^{12}\,\mathrm{GeV}\) \cite{Bansal:2024uzr}, exceeding the BBN lower bound by many orders of magnitude.\vspace*{2pt}

\textbf{CMB anisotropies} \textit{(constrain $  n_s  $, $  r  $, and $  \Delta N_{\rm eff}  $)}: As the universe cools below $T\sim 0.3\,\mathrm{eV}$, photons decouple from the baryon-photon fluid at redshift $z_*\approx 1090$, giving rise to the Cosmic Microwave Background (CMB). The Planck satellite has measured the temperature and polarisation  anisotropies of the CMB and extracted the corresponding angular power spectra. These spectra encode the primordial scalar and tensor perturbations ($  \tilde{A}_s  $, $  n_s  $, $  r  $) after they have been processed by the photon-baryon fluid prior to decoupling, together with any extra relativistic degrees of freedom parametrised by $\Delta N_{\rm eff}$.
Theoretical angular power spectra computed within $\Lambda$CDM and its extensions are then compared with the observed spectra to constrain the cosmological parameters. The predicted values of $  n_s  $, $  r  $, and $  \Delta N_{\rm eff}  $ in our three scenarios (Table \ref{pheno-table}) are confronted with the latest observational results below.

{\setlength{\parindent}{17pt}The most comprehensive and up-to-date constraint on $n_s$ in base $\Lambda$CDM which has no extra dark radiation, combines all CMB and BAO data from SPT, Planck, ACT, and BICEP/Keck \cite{Balkenhol:2025wms}, giving:}
\vspace{-6pt}
\begin{equation}
    n_s = 0.9728 \pm 0.0029 \quad \text{at 68\% CL for } \Delta N_{\text{eff}} = 0 \quad \text{(SPA+BK+DESI).}\vspace{-6pt}
\end{equation}
This \looseness=-1 is comparable with CMB-SPA+lensing+DESI $n_s = 0.9730 \pm 0.0026$ \cite{McDonough:2025lzo}.\newpage

Scenario I has vanishing extra dark radiation. Its predicted value of
\vspace{-6pt}
\begin{equation}
    n_s \simeq 0.9765 \quad \text{lies within } 1.28\sigma \text{ of the SPA+BK+DESI result,}\vspace{-6pt}
\end{equation}
and $1.35\sigma$ of \looseness=-2 the CMB-SPA+lensing+DESI result, notable improvements over the $2.14\sigma$ deviation from the Planck PR4 TTTEEE+lensing+BAO $n_s=0.9690\pm 0.0035$ \cite{Tristram:2023haj}. The CMB-only constraint from \cite{Balkenhol:2025wms},\vspace{-6pt}
\begin{equation}
    n_s = 0.9682 \pm 0.0032 \quad \text{at 68\% CL for } \Delta N_{\text{eff}} = 0 \quad \text{(SPA+BK),}\vspace{-3pt}
\end{equation}
gives a $2.59\sigma$ deviation, $0.33\sigma$ higher than the $2.26\sigma$ obtained using the Planck PR4 TTTEEE +lensing $n_s = 0.9679 \pm 0.0038$ \cite{Tristram:2023haj}. The difference between the CMB+BAO and CMB-only constraints arises from a ${\sim}\,2.8\sigma$ tension between DESI BAO data and CMB data, whose origin is under active investigation \cite{McDonough:2025lzo,Balkenhol:2025wms}.

A similar pattern is seen in the ACT dataset combinations: the CMB+BAO combination ACT+DESI \cite{McDonough:2025lzo} yields {\it near-perfect agreement of $\mathit{-0.03\sigma}$} with $n_s = 0.9737 \pm 0.0025$, P-ACT-LB2 \cite{AtacamaCosmologyTelescope:2025blo} yields a very good agreement of $0.43\sigma$ with $n_s = 0.9752 \pm 0.0030$, while the CMB-only combination P-ACT+lensing \cite{McDonough:2025lzo} yields a deviation of $1.44\sigma$ with $0.9713 \pm 0.0036$.

A brief check shows that including subleading loop corrections further improves agreement.

\enlargethispage{1\baselineskip}
\begin{table}[h]
\centering
\begin{tabular}{lccccc}
\toprule
\textbf{Scenario} & $T_{\rm rh}$ (GeV) & $N_e$  & $\Delta N_{\rm eff}$ & $n_s$ & r  \\
\midrule
I (SM on D7s, $\tau_\phi$ wrapped) & $4 \times 10^{10}$ & 53  & $\simeq 0$ & 0.9765 & $1.7 \times 10^{-5}$\\
II (SM on D7s, $\tau_\phi$ not wrapped) & $3 \times 10^{12}$ & 52  & $\simeq 0.14$ & 0.9761 & $1.7 \times 10^{-5}$\\
{III (SM on D3s)\footnotemark} & $2 \times 10^8$ & 51.6 &$\simeq 0.09$ & 0.9758 & $1.8 \times 10^{-5}$\\
\bottomrule
\end{tabular}
\vspace{-7pt}
\caption{Post-inflationary parameters and observational predictions for the three scenarios.}
\label{pheno-table}
\vspace{-8pt}
\end{table}
\footnotetext{Scenarios IIIa and IIIb have identical late-time parameters and are grouped as Scenario III. We adopt $Z = 4$ to satisfy the ACT DR6 constraint $Z \gtrsim 2.9$ derived from $\Delta N_{\rm eff} < 0.17$ at 95\% CL \cite{AtacamaCosmologyTelescope:2025nti}, updating the earlier choice $Z = 2$ in \cite{Bansal:2024uzr}.}

For \looseness=-1 Scenarios II and III, where $\Delta N_{\rm eff} \simeq 0.14$ and $\simeq 0.09$ respectively, a dedicated fit with $\Delta N_{\rm eff}$ fixed at the relevant value would be needed for a precise comparison of $n_s$. However, non-zero $\Delta N_{\rm eff}$ is known to increase the best-fit value of $n_s$ extracted from CMB(+BAO) data, which is expected to improve agreement with our predictions. Both scenarios satisfy the one-sided ACT DR6 bound $\Delta N_{\rm eff} < 0.17$ at 95\% CL \cite{AtacamaCosmologyTelescope:2025nti}. The full $\Lambda$CDM + $N_{\rm eff}$ fit gives $N_{\rm eff} = 2.86 \pm 0.13$ at 68\% CL, corresponding to $\Delta N_{\rm eff} = -0.18 \pm 0.13$, with the mildly negative central value driven by the ACT DR6 data preferring slightly less damping. Scenario I with $\Delta N_{\rm eff} \simeq 0$ is fully compatible with this measurement. Scenarios II and III satisfy the one-sided bound but lie in the upper tail of the two-sided posterior. Future CMB data will clarify whether the mildly negative central value of $\Delta N_{\rm eff}$ persists.

In all the three scenarios the model yields a tensor-to-scalar ratio of order\vspace{-8pt}
\begin{equation}\label{rPrediction}
r\simeq 2\times 10^{-5}\,, 
\vspace{-9pt}
\end{equation}
which comfortably satisfies the current upper bound  $r < 0.034$ at 95\% CL \cite{Balkenhol:2025wms} and is roughly five orders of magnitude larger than the $r\sim 10^{-10}$ prediction of the original non-perturbative blow-up inflation model.

\enlargethispage{1\baselineskip}
\section{Subleading Loop Corrections}

The loop correction in eqs. \eqref{eq:leading-f} - \eqref{deltaV_loop} was derived under the assumption that the blow-up cycle $\tau_\phi$ is small enough for the leading term in the expansion of the loop correction function $f$ to dominate. This corresponds to the regime $\tau_\phi \ll \mathcal{V}^{2/3}$ ($\phi \ll 1$). As $\phi \to \mathcal{O}(1)$, subleading terms in the expansion of $f$ become relevant:\vspace{-7pt}
\begin{equation}
    f \simeq \frac{\mathcal{V}^{1/3}}{\sqrt{\tau_\phi}}\left(1 + \frac{\sqrt{\tau_\phi}}{\mathcal{V}^{1/3}} + \frac{\tau_\phi}{\mathcal{V}^{2/3}}+\dots \right),\vspace{-6pt}
\end{equation}
with each successive term suppressed by $\sqrt{\tau_\phi}/\mathcal{V}^{1/3} \sim \phi^{2/3}$. This modifies the inflationary potential to:\vspace{-6pt}
\begin{equation}
    V=V_0\left(1 -c_{\rm loop}\, b \left[\frac{1}{\phi^{2/3}} + \mathfrak{a} + \mathfrak{b} ~\phi^{2/3} +\dots\right]\right),\vspace{-2pt}
\end{equation}
where $\mathfrak{a}, \mathfrak{b} \sim \mathcal{O}(1)$. The constant $\mathfrak{a}$ negligibly shifts the plateau height and can be dropped. The constant $\mathfrak{b}$ genuinely corrects the potential shape, and depending on its sign and value, new classes of slow-roll inflationary models could emerge. Focusing on the leading correction, the number of e-foldings becomes:\vspace{-8pt}
\begin{equation}
    N_e\simeq \frac{9}{16} \frac{\phi_*^{8/3}}{b\,c_{\rm loop}} (1 + 2\,\mathfrak{b}\, \phi_*^{4/3}),\vspace{-2pt}
\end{equation}
and solving for $\mathcal{V}$ using the scalar perturbation amplitude gives:\vspace{-6pt}
\begin{equation}
    \mathcal{V}=\frac{\mathcal{A}}{\phi^{8}_*} \left(1+2\,\mathfrak{b}\, \phi^{4/3}_*\right)^{-3}.\vspace{-6pt}
\end{equation}
For $\mathfrak{b} > 0$, both $\phi_*$ and $\mathcal{V}$ are reduced. This means that  a predetermined number of e-foldings and slow-roll can be achieved without pushing $\phi_*$ to large values, pulling the inflationary trajectory further into the Kähler cone interior. This improves theoretical control and provides a posteriori justification for the simplest model where the $\mathfrak{b}$-correction was neglected. Determining the signs and magnitudes of $\mathfrak{b}$ and higher coefficients requires explicit loop calculations on a specific Calabi-Yau geometry, which we leave for future work.

\enlargethispage{1\baselineskip}
\section{Classification of \Kahler\ Moduli Inflation Models}

The general structure of the potential of all LVS Kähler moduli inflation models is:\vspace{-6pt}
\begin{equation}\label{V_tot}
V_{\rm tot} (\mathcal{V},\tau_\phi)= V_{\rm lead}(\mathcal{V}) -  V_{\rm sub}(\mathcal{V},\tau_\phi)\,\vspace{-4pt}
\end{equation}
with $V_{\rm sub}(\mathcal{V},\tau_\phi) \ll V_{\rm lead}(\mathcal{V})$. $V_{\rm tot}$ is stabilised at $\vol\!=\!\langle\mathcal{V}\rangle$ and $\tau_\phi\!=\!\langle\tau_\phi\rangle$ where $\langle\mathcal{V}\rangle$ and $\langle\tau_\phi\rangle$ denote the vacuum expectation values of the volume modulus and the inflaton respectively. This sets $V_{\rm tot}(\langle\mathcal{V}\rangle,\langle\tau_\phi\rangle)=0$, which causes eq. \eqref{V_tot} to give\vspace{-4pt}
\begin{equation}
    V_{\rm lead}(\langle\mathcal{V}\rangle) = V_{\rm sub}(\langle\mathcal{V}\rangle,\langle\tau_\phi\rangle)\,.\vspace{-4pt}
\end{equation}
The inflationary potential, which is supposed to be stabilised with respect to $\vol$, is $V_{\rm tot} (\langle\mathcal{V}\rangle,\tau_\phi)$.\vspace{-4pt}
\begin{equation}
V_{\rm tot} (\langle\mathcal{V}\rangle,\tau_\phi) = V_{\rm sub}(\langle\mathcal{V}\rangle,\langle\tau_\phi\rangle)
 - V_{\rm sub}(\langle\mathcal{V}\rangle,\tau_\phi) = V_{\rm sub}(\langle\mathcal{V}\rangle,\langle\tau_\phi\rangle) \left[1 - \frac{V_{\rm sub}(\langle\mathcal{V}\rangle,\tau_\phi)}{V_{\rm sub}(\langle\mathcal{V}\rangle,\langle\tau_\phi\rangle)}\right].\vspace{-4pt}
\end{equation}
This potential takes the typical plateau-like form of K\"ahler moduli inflation:\vspace{-6pt}
\begin{equation}
V = V_0\left[1 - g(\phi)\right],\vspace{-6pt}
\end{equation}
with\vspace{-6pt}
\begin{equation}
V_0\equiv V_{\rm sub}(\langle\mathcal{V}\rangle,\langle\tau_\phi\rangle)   \qquad\text{and} \qquad g(\phi)\equiv \frac{V_{\rm sub}(\langle\mathcal{V}\rangle,\tau_\phi(\phi))}{V_{\rm sub}(\langle\mathcal{V}\rangle,\langle\tau_\phi\rangle)}\,.
\end{equation}
The functional form of $g(\phi)$ depends on two features:\vspace{-6pt}
\begin{itemize}[itemsep=-1em]
    \item[1)] The first is the origin of the subleading potential $V_{\rm sub}(\langle\mathcal{V}\rangle,\tau_\phi)$ -- either non-perturbative or perturbative: \vspace{-6pt}
    \begin{itemize}
    \item[(i)] Non-perturbative effects (exponentially suppressed):\vspace{-6pt}
\begin{equation}
V_{\rm sub}(\langle\mathcal{V}\rangle,\tau_\phi)\propto e^{-k\tau_\phi} \underset{\tau_\phi\to\infty}{\longrightarrow} 0\qquad\text{for}\quad k>0\,,\vspace{-10pt}
\end{equation}
    \item[(ii)] Perturbative effects (typically power-law):\vspace{-7pt}
\begin{equation}
V_{\rm sub}(\langle\mathcal{V}\rangle,\tau_\phi)\propto \frac{1}{\tau_\phi^p} \underset{\tau_\phi\to\infty}{\longrightarrow} 0 \qquad\text{for}\quad p>0\,.
\end{equation}
    \end{itemize}
    \item[2)] The second is the topology of $\tau_\phi$, which determines the relation between $\tau_\phi$ and the canonical inflaton $\phi$: \vspace{-6pt}
\begin{itemize}
\item[(iii)] Bulk fibre modulus (exponential relation): \vspace{-6pt}
\begin{equation}
    \tau_\phi = e^{\lambda \phi}\qquad\text{with}\qquad \lambda \sim \mathcal{O}(1)\,,\vspace{-6pt}
\end{equation}
\item[(iv)] Local blow-up modulus (power-law relation):\vspace{-6pt}
\begin{equation}
\tau_\phi = \mu\,\mathcal{V}^{2/3}\,\phi^{4/3} \qquad\text{with}\qquad \mu\sim\mathcal{O}(1)\,.\vspace{-6pt}
\end{equation}
\end{itemize}
\end{itemize}
The four resulting classes are shown in Table~\ref{tab:classification}.\enlargethispage{1\baselineskip}

\begingroup
\setlength{\tabcolsep}{2.5pt} 
\begin{table}[h]
\centering
\begin{tabular}{lccc}
\toprule
\textbf{Effects} & \textbf{Bulk fibre modulus} & \textbf{Local blow-up modulus} \\
\midrule
Non-perturbative & Fibre Inflation: $\!g \!\propto\! e^{-k e^{\lambda\phi}}$ & Blow-up Inflation: $\!g\!\propto\! e^{-k\mu(\mathcal{V}\phi^{2})^{2/3}}\!$ {\small (\textit{unviable})\footnotemark} \\
Perturbative & Loop Fibre Inflation: $\!g \!\propto\! e^{-p\lambda\phi}$ & \textbf{Loop Blow-up Inflation:} $\!g \!\propto\! (\vol\phi^2)^{-2p/3}$ \\
\bottomrule
\end{tabular}
\caption{Classification of \Kahler\ moduli inflation models. Loop Blow-up Inflation is the first example with a power-law potential within LVS Kähler moduli inflation.}
\label{tab:classification}
\vspace{-6pt}
\end{table}
\endgroup
\footnotetext{The original non-perturbative blow-up inflation model is rendered physically unviable by unavoidable string loop corrections, as discussed in sec. \ref{inev_corr}.}
Loop Blow-up Inflation represents the first example in this class of constructions with a power-law inflationary potential, in contrast to the exponential potentials of all previous models. In all models except Loop Blow-up Inflation, the inflationary plateau condition $g(\phi) \ll 1$ is typically satisfied in the large-field regime; in the latter, it requires $\phi \lesssim 1$.

\enlargethispage{1\baselineskip}
\section{Conclusions and Outlook}

We have presented Loop Blow-up Inflation, a new class of string inflation models where the inflationary potential is generated by string loop corrections to the \Kahler\ potential. We found that these corrections are unavoidable in blow-up inflation models and change the original model in the following ways:
\vspace*{-5pt}
\begin{itemize}
[itemsep=-0.4em,leftmargin=1em]
\item \textbf{Shift the slow-roll field range.} The original model has slow-roll near the non-perturbative minimum at small $\phi$ (where $a_\phi\tau_\phi \sim \ln\mathcal{V}$). Loop corrections invalidate slow-roll there and create a new slow-roll region at larger $\phi \lesssim 1$.
\item \textbf{Change the potential from exponential to power-law.} The exponential plateau $V = V_0(1 - \kappa_1 e^{-\kappa_2\phi})$ of the original model is replaced by a power-law plateau $V = V_0(1 - b\,c_{\rm loop}/\phi^{2/3})$, with slow-roll guaranteed by the smallness of $b\,c_{\rm loop}$. Loop Blow-up Inflation is the first Kähler moduli inflation model with a power-law potential.
\item \textbf{Change the cosmological predictions.} The tensor-to-scalar ratio increases from $r \sim 10^{-10}$ in the original model to $r \sim 2 \times 10^{-5}$, roughly five orders of magnitude larger. Other cosmological predictions also change.
\end{itemize}\vspace*{-3mm}
Including subleading loop corrections improves the Loop Blow-up Inflation model by reducing the required $\phi_*$, making it more robust.

Promising directions for future work include a first-principles determination of the loop coefficient $c_{\rm loop}$, whose sign and magnitude critically impact the viability of the model. This may be feasible in simple blow-up geometries such as the blown-up $\mathbb{C}^3/\mathbb{Z}_3$, where an explicit Ricci-flat metric is known. It would also be interesting to include additional perturbative corrections such as higher F-term $\alpha'^3$ effects, and to explicitly compute subleading loop corrections in concrete Calabi-Yau geometries, especially near Kähler cone boundaries.
\vspace{16pt}

\section*{Acknowledgments}

This proceedings contribution provides an overview of Loop Blow-up Inflation in light of the latest CMB and BAO observations. It is based on work originally published in \cite{Bansal:2024uzr}, done in collaboration with L.~Brunelli, M.~Cicoli, A.~Hebecker, and R.~Kuespert. In addition to summarising the original findings, this work updates the cosmological predictions and their comparison with recent observations. For the vanishing extra dark radiation scenario, predictions are compared with the latest combined analyses of ACT, DESI, SPT, BICEP/Keck, and Planck, while scenarios with non-zero extra dark radiation are assessed using ACT DR6 constraints. Scenario III is also revised using an updated Giudice-Masiero coefficient required by tighter bounds on extra dark radiation, and new comparisons with current $n_s$ and $\Delta N_{\text{eff}}$ measurements are presented.

The author is grateful to the collaborators on the original work, and additionally thanks Michele Cicoli and Arthur Hebecker for their helpful input on this manuscript. The author also extends gratitude to the organisers of the 2025 Workshop on Quantum Gravity and Strings at the Corfu Summer Institute for the opportunity to present this work and for a stimulating meeting. Additionally, the author thanks Kurt Weninger for facilitating and encouraging participation in the workshop.

\bibliographystyle{JHEP}
{
\setlength{\baselineskip}{1.13em}
\bibliography{references}}

\end{document}